# Symmetry-protected photonic chiral spin textures by spin-orbit coupling

*Peng Shi\*, Luping Du\*\*, Mingjie Li, and Xiaocong Yuan\*\*\**

Nanophotonics Research Centre, Shenzhen Key Laboratory of Micro-scale Optical Information Technology, Institute of micro/nano optoelectronics, Shenzhen University, Shenzhen, 518060, China.

\*shipeng@szu.edu.cn.
\*\*lpdu@szu.edu.cn.
\*\*\*xcyuan@szu.edu.cn.

Keywords: spin-orbit interaction, chiral spin textures, skyrmion, symmetry, conservative quantity; variational method

**Abstract**: Chiral spin textures are researched widely in condensed matter systems and show potential for spintronics and storage applications. Along with extensive condensed-matter studies of chiral spin textures, photonic counterparts of these textures have been observed in various optical systems with broken inversion symmetry. Unfortunately, the resemblances are only phenomenological. This work proposes a theoretical framework based on the variational theorem to show that the formation of photonic chiral spin textures in an optical interface is derived from the system's symmetry and relativity. Analysis of the optical system's rotational symmetry indicates that conservation of the total angular momentum is preserved from the local variations of spin vectors. Specifically, although the integral spin momentum does not carry net energy, the local spin momentum distribution, which determines the local subluminal energy transport and minimization variation of the square of total angular momentum, results in the chiral twisting of the spin vectors. The findings here deepen the understanding of the symmetries, conservative laws and energy transportation in optical system, construct the comparability in the formation mechanisms and geometries of photonic and condensed-matter chiral spin textures, and suggest applications to optical manipulation and chiral photonics.

**1. Introduction**

Recently, the photonic analogy of topological physics has become a rapidly-developing research area in which the geometrical and topological concepts can be exploited to control the behavior of light.[1] By engineering the 'extrinsic' spin-orbit interactions in artificial photonic structures with broken time-reversal symmetry, a variety of fascinating phenomena and chiral spin textures have been demonstrated in momentum-space that offer potential for application to fields including integrated optoelectronic devices and quantum technologies.[2-5] On the other hand, owing to the 'intrinsic' spin-orbit coupling that is governed by Maxwell's theory, many topology-like phenomena in the real space have been reported, including polarization Möbius strips,[6, 7] polarization vortex,[8, 9] along with various chiral textures such as optical domain walls,[10] photonic skyrmions and merons.[11-19] Among these phenomena, the magnetic skyrmion,[20-22] which is a topologically non-trivial spin whirling after the nuclear physicist Tony Skyrme, can be formed via the spin-orbit interaction (Dzyaloshinskii-Moriya interaction: DMI) in an electronic system lack of inversion symmetry and minimizing the magnetic energy cost. Photonic skyrmions, which have chiral spin textures that can be considered as the optical manifestation of magnetic skyrmions, were discovered and have attracted widespread interest in the spin optics, chiral quantum optics and photoelectric interaction fields.[22-30]

Optical spin-orbit interactions play a significant role in the formation of photonic chiral spin textures.[31, 32] Light has two fundamental dynamical properties: momentum and angular momentum (AM),[31-34] in which the spin angular momentum (SAM) is associated with the polarization ellipticity and the orbital angular momentum (OAM) is determined by either the vortex phase or the optical beam trajectory.[33] The interplay between these two angular momenta has been studied extensively over the past few years and has been shown to result in many fascinating phenomena.[34] Using the Dirac-like form of Maxwell's equations,[35] one can draw a visual picture of the spin-orbit interactions and topological properties between photons and electrons. For electrons in topological insulators, the electric potential of the Dirac equation in the nonrelativistic limit leads to strong spin-orbit interactions that cause the spin, momentum, Coulomb interaction and external electric fields to couple together, resulting in the emergence of the topologically nontrivial band structures.[36] In the optical system, the optical potential is determined exclusively by the spatial distributions of the relative permittivities and permeabilities of the materials. The spatial variance properties of optical materials can also lead to the optical spin-orbit interactions, particularly in the optical near-field, because the spin-orbit interactions become conspicuous at subwavelength scales and show intrinsic coupling between the SAM and the position of light.[37-39] In addition, there is a solution of bound states, which are quite robust against the interface roughness or other factors, at the interface between two media with positive and negative mass or energy gap from the Dirac equation. Whereas for photons, there also exists the surface mode at the interface, e.g. evanescent plane wave.[40] The evanescent plane wave is topological nontrivial ($\mathbb{Z}_4$ invariant) and possesses a feature of spin-momentum locking.[31, 41, 42] However, comparing with the scalar electronic wave function,[43] the vector feature of the photon wave function gives rise to the chiral spin textures that are analogous to textures of the topological excitations.[20-22, 44] To date, this resemblance is only in phenomenological terms and most of the relevant researches has focused on the experimental characterization and applications of the photonic chiral spin textures;[45-48] the formation and stability properties of these textures have yet to be clarified in comparison to those of the condensed state system.

Here, we propose a theoretical framework to explain the formation of photonic chiral spin textures using the symmetry of the optical system. Any differentiable symmetry in the action of a physical system has a corresponding conservation law from Noether's first theorem,[49] which states that the rotational symmetry is connected to the conservation of total AM and the translational symmetry is related to the conservation of momentum. By employing the variational method and performing the same procedure as the magnetic skyrmion (i.e. translational symmetry in the outer normal direction), we determine the skyrmion-like chiral spin textures in the cylindrical coordinates and the domain wall-like chiral spin textures in the Cartesian coordinates, which indicate the formation of chiral spin textures stems from the conservation of the total AM and that its stability is protected by the symmetry. The result is general and only demands the conditions of symmetry and subluminal energy transportation imposed by special relativity. Therefore, this result can be extended flexibly to electromagnetic (EM) waves from the radio-frequency (RF) to ultraviolet (UV) spectral ranges in other coordinate systems, including both parabolic and ellipsoidal coordinates, and also to various classical waves, including acoustic, fluid, elastic and gravitational waves. Our findings are important for gaining an understanding of the chiral spin textures and spin dynamics in classical waves and for applications including optical manipulation, chiral imaging, nanometrology and prospective on-chip optoelectronic technologies.

**2. Symmetry-induced chiral spin textures in various coordinate systems**

The photonic chiral spin texture, which is an optical counterpart of the topological defects in condensed matter system, can be constructed using *p*-polarized or *s*-polarized surface EM wave, such as: evanescent waves,[50] Bloch surface waves,[51] and surface plasmon polaritons (SPPs),[11] at a vacuum/ isotropous material interface with broken inverse symmetry.[31] Most of structures sustaining the surface waves are lossless. Even for the SPP configuration, the loss mechanism does not affect the symmetry of optical system and it can be ignored by only considering the probability distribution of a single EM wave packet at the interface.[52, 53] Moreover, from the former experimental results related to the photonic chiral textures,[11, 12, 17, 18, 31, 45-48] one can find that the material loss primarily influences the intensity (photon number) but not the orientation of vector direction. Thus, we can ignore the material losses here for convenience, as the former researches.[41, 50] To describe the spin-orbit interactions and conservative properties of the general photonic chiral spin textures, we introduce the momentum and the AM of light at optical interface using the optical Dirac equation:[35]

$$\hat{\mathbf{H}}|\psi_0\rangle = c\hat{\boldsymbol{\tau}} \cdot \hat{\mathbf{p}}|\psi_0\rangle + \hat{\boldsymbol{\gamma}}\mathbf{V}|\psi_0\rangle = i\hbar\frac{\partial}{\partial t}|\psi_0\rangle, \qquad (1)$$

where $\hat{\mathbf{H}} = c\hat{\boldsymbol{\tau}} \cdot \hat{\mathbf{p}} + \hat{\boldsymbol{\gamma}}\mathbf{V}$ is the photonic Hamilton operator with $c$ the velocity of light in vacuum; $|\psi_0\rangle = [\sqrt{\varepsilon_0}\mathbf{E}; i\sqrt{\mu_0}\mathbf{H}]/2$ is the position-dependent 6-vector photon wave function formed by electric and magnetic fields; $\hat{\boldsymbol{\tau}}_i$ ($i$=1,2,3) and $\hat{\boldsymbol{\gamma}}$=[**I**,**0**;**0**,-**I**] represent the four Dirac matrices and satisfy [$\hat{\boldsymbol{\tau}}_i$,$\hat{\boldsymbol{\gamma}}$]=0, where the $\hat{\boldsymbol{\tau}}_i$=[**0**,$\hat{\mathbf{S}}_i$;$\hat{\mathbf{S}}_i$,**0**] and $\hat{\mathbf{S}}_i$ represents the spin-1 matrix in SO(3); $\hat{\mathbf{p}} = -i\hbar\nabla$ is the canonical momentum operator with $\hbar$ the reduced Plank constant; V=$\hbar\omega$[(1−$\varepsilon_r$)**I**, **0**; **0**, ($\mu_r$−1)**I**] is the optical potential induced by the material properties, where $\omega$ the angular frequency of light. Here, the symbols **I** and **0** denote the 3×3 unit matrix and the null matrix, respectively. From **Equation 1**, in the vacuum half-space, the kinetic momentum density (MD) can be defined as **p**=⟨$\psi_0$|$\hat{\boldsymbol{\tau}}$/$c$|$\psi_0$⟩.[31, 32, 35] In general, the kinetic MD can be decomposed into the contributions of orbit and spin components (**p**=**p**$_o$+**p**$_s$): the orbit (canonical) MD **p**$_o$=⟨$\psi_0$|$\hat{\mathbf{p}}$|$\psi_0$⟩/$\hbar\omega$ and spin MD **p**$_s$=∇×⟨$\psi_0$|$\hat{\boldsymbol{\Sigma}}$|$\psi_0$⟩/2$\hbar\omega$, in which $\hat{\boldsymbol{\Sigma}}$=$\hbar$[$\hat{\mathbf{S}}$,**0**;**0**,$\hat{\mathbf{S}}$] is the spin operator of spin-1 photon.[35, 50]

Firstly, we investigate the angular momenta of photonic chiral spin textures and their conservative properties. The intrinsic spin angular momentum (iSAM) density of a single wave packet is expressed as **Σ**=⟨$\psi_0$|$\hat{\boldsymbol{\Sigma}}$|$\psi_0$⟩/$\hbar\omega$,[35, 50] whereas the total angular momentum density of light can be expressed as **J**=**r**×**p**=**L**+**S**, where **L**=**r**×**p**$_o$ and **S**=**r**×**p**$_s$ are the orbital and spin AM densities, respectively, with position vector **r**. For the photonic chiral spin textures at an optical interface, the total AM is conservative because $\partial \hat{\mathbf{J}}/\partial t = i[\hat{\mathbf{H}},\hat{\mathbf{J}}]/\hbar = 0$, where $\hat{\mathbf{J}}$ is the total AM operator (see Supplemental Text I), whereas the spin and orbital AM densities are not conservative individually because of the mutual coupling and conversion of these two types of angular momenta induced by optical potential (Generally, the spin and orbital AMs of nonparaxial beams are not conservative individually owing to the spin-orbit coupling, while in a paraxial system that the spin-orbit coupling can be ignored, the spin and orbital AMs can be regarded as be conservative individually; here, the nonparaxial system is constructed with special optical potential, which introduces the strong spin-orbit interaction in the optical system with interface). This is logical because the optical system has the rotational symmetry along the normal direction. The spin-orbit conversion behavior can be understood with the aid of the integral forms of the AM densities: $\langle \mathbf{J} \rangle = \langle \mathbf{r} \times \mathbf{p} \rangle$, $\langle \mathbf{L} \rangle = \langle \mathbf{r} \times \mathbf{p}_o \rangle$ and $\langle \mathbf{S} \rangle = \langle \mathbf{r} \times \mathbf{p}_s \rangle$,[50] where the Dirac notation corresponds to the two-dimensional integral of the physical quantities on the horizontal plane oriented parallel to the optical interface. Therein, the orbital AM consists of extrinsic and intrinsic parts, given by $\langle \mathbf{L} \rangle^{\text{ext}} = \langle \mathbf{r} \rangle \times \langle \mathbf{p}_o \rangle$ and $\langle \mathbf{L} \rangle^{\text{int}} = \langle \mathbf{L} \rangle - \langle \mathbf{L} \rangle^{\text{ext}}$, respectively, where $\langle \mathbf{r} \rangle$ is the centroid of the beam. When the stokes theorem is used and it is assumed that the iSAM disappears at infinity in a physical sense, the integral spin MD over the entire plane $\Omega$ oriented parallel to the optical interface vanishes because

$$\langle \mathbf{p}_s \rangle = \int_\infty \mathbf{p}_s d\Omega = \int_\infty \frac{1}{2} \nabla \times \mathbf{\Sigma} d\Omega = \frac{1}{2} \oint_\infty \mathbf{\Sigma} dC = 0. \tag{2}$$

Therefore, the SAM is intrinsic because $\langle \mathbf{S} \rangle^{\text{ext}} = \langle \mathbf{r} \rangle \times \langle \mathbf{p}_s \rangle = 0$, and thus it follows that $\langle \mathbf{S} \rangle^{\text{int}} = \langle \mathbf{\Sigma} \rangle$. For the photonic chiral spin textures in the vacuum half-space, the following relationship has been determined between the iSAM and the kinetic MD: $\mathbf{\Sigma} = \nabla \times \mathbf{p}/2k_0^2$, where $k_0$ is the wave vector in vacuum.[31] In addition, because $\langle \mathbf{\Sigma} \rangle = \int_\infty \mathbf{\Sigma} d\Omega = \oint_\infty \mathbf{p}/2k_0^2 dC = 0$ due to the MD, which is dependent on $1/r$ for the cylindrical wave, disappears at infinity, the integral of iSAM $\langle \mathbf{\Sigma} \rangle$ also vanishes. Therefore, we can determine that $\langle \mathbf{J} \rangle = \langle \mathbf{L} \rangle^{\text{int}}$, which indicates that the total AM is conservative across the optical interface and that there is spin-to-orbit conversion for photonic chiral spin textures. For example, the photonic skyrmion can be excited by focusing a circularly polarized plane wave with a longitudinal helix σ that carries an additional vortex phase with a topological charge *m*, and the quantum number of Bessel-type surface EM wave is $l=\sigma+m$ (see Supplemental Text III).

In addition, the square modulus of total AM is also conservative because $\partial \hat{\mathbf{J}}^2/\partial t=0$ (see Supplemental Text I). Using the photonic skyrmion as an example, it can be deduced that $\langle \mathbf{J}_i \cdot \mathbf{J}_i \rangle = (\sigma+m)^2 \hbar^2$ for the square modulus of total AM in incident wave, which results in the photonic skyrmion has the feature that $\langle \mathbf{J}_{SK} \cdot \mathbf{J}_{SK} \rangle = \langle \mathbf{J}_i \cdot \mathbf{J}_i \rangle = l^2 \hbar^2$ because of the conservation of square modulus of total AM. It should be noted here that the square modulus of total AM can be decomposed into $\langle \mathbf{J}_{SK} \cdot \mathbf{J}_{SK} \rangle = \langle \mathbf{L} \cdot \mathbf{L} + 2\mathbf{L} \cdot \mathbf{S} + \mathbf{S} \cdot \mathbf{S} \rangle$, in which there is a spin-orbit coupling term that is analogous to the corresponding term in quantum physics.[54]

It is well known that, in condensed state physics, the formation of a magnetic skyrmion is due to the DMI and the minimizing of magnetic energy cost (see Supplemental Text IV). Here, we construct a similar theory with conservation of total AM. Therefore, the physical quantity $\langle \mathbf{J}_{SK} \cdot \mathbf{J}_{SK} \rangle$ would be a candidate in determining the formation of the photonic chiral spin textures. Because the square modulus of total AM $\langle \mathbf{J}_{SK} \cdot \mathbf{J}_{SK} \rangle$ is conservative in this optical system, any change of the local distribution of iSAM would break this conservation and thus the variation of the square modulus of the total AM on the orientation of the local spin vector must vanish, i.e. $\delta \langle \mathbf{J}_{SK} \cdot \mathbf{J}_{SK} \rangle = 0$. From simple derivations (see Supplemental Texts II and VI), the variation of the square modulus of the total AM can be converted into

$$\delta \langle \mathbf{J}_{SK} \cdot \mathbf{J}_{SK} \rangle = \frac{2l}{\omega} \mathbf{W} \cdot \langle \delta \mathbf{S} \rangle + \delta \langle \mathbf{S} \cdot \mathbf{S} \rangle$$
$$= \delta \langle r^2 \mathbf{p}_s \cdot \mathbf{p}_s \rangle = \delta \left\langle r^2 \left\{ -\mathbf{\Sigma} \cdot \nabla_{\perp}^2 \mathbf{\Sigma} - \mathbf{\Sigma} \cdot \frac{\partial^2}{\partial \mathrm{n}^2} \mathbf{\Sigma} + \nabla \cdot \left[ \mathbf{\Sigma} \times (\nabla \times \mathbf{\Sigma}) \right] \right\} \right\rangle, \quad (3)$$

The three terms in the second line of **Equation 3** above can be regarded as the photonic exchange energy term, the anisotropy term induced by the optical interface and the DMI term, respectively. The symbol *n* here denotes the outer normal direction of interface and $\nabla_\perp^2$ is the transverse Laplace operator. It should be noted here that we select that energy density $W=\langle\psi_0|\psi_0\rangle=1$ for a single wave packet.

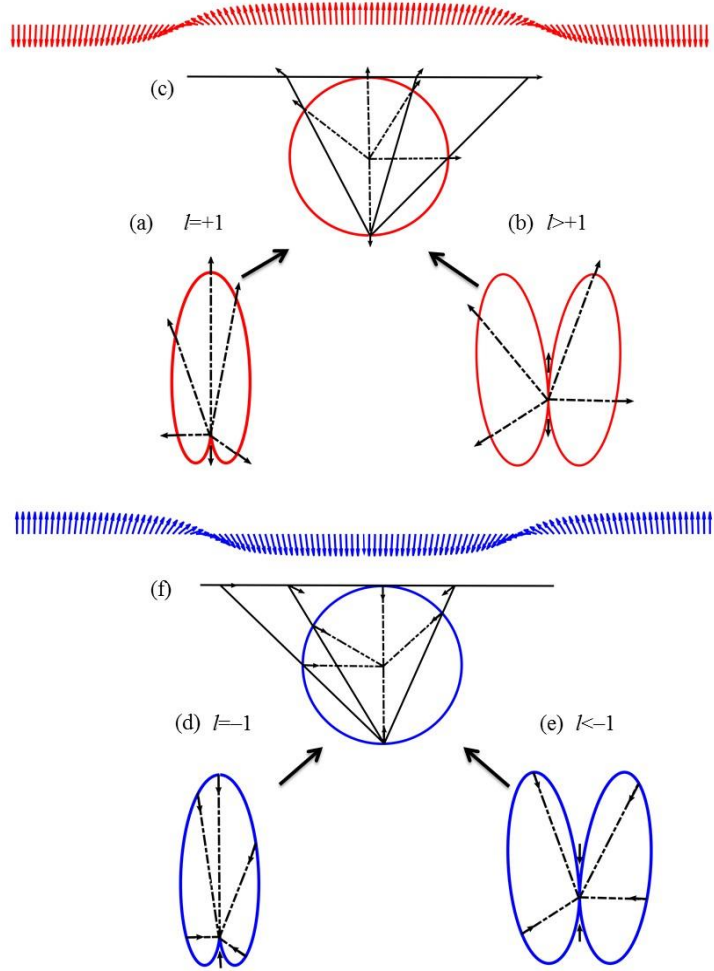

**Figure 1.** Geometry of photonic skyrmions. Projections of the skyrmion-like spin textures in cylindrical coordinates with quantum numbers (a) *l*=+1 and (b) *l*>+1 to a specific spherical coordinate; (c) corresponding projection of the normalized spin texture (top panel) to a specific spherical coordinate. These projections indicate that the skyrmion numbers of the spin textures for photonic skyrmions with positive integer quantum numbers are -1 uniformly. The directional vector of the iSAM is in the outer direction along the connecting line between the centre (0, 0) and red rings and its relative intensity is proportional to the length of connecting line. Projection of the skyrmion-like spin texture in cylindrical coordinates with quantum numbers (d) *l*=-1 and (e) *l*<-1 to a specific spherical coordinate; (f) corresponding projection of the normalized spin texture (top panel) to a specific spherical coordinate. These projections indicate the skyrmion numbers of the spin textures for photonic skyrmions with negative integer quantum numbers are +1. The directional vector of iSAM is in the inner direction along the connecting line between the centre (0, 0) and blue rings. It should be noted that the geometries of photonic skyrmions with quantum numbers *l*>+1 or *l*<-1 are approximately identical. Therefore, only (b) and (e) are exhibited here. The skyrmion-like spin texture is cylindrically symmetrical and thus the 2D contours in the *xz*-plane are shown here. The photonic skyrmions are constructed by the *p*-polarized light at air-gold interface.

Subsequently, we present the formation of the skyrmion-like chiral spin texture in cylindrical coordinates ($r$, $\varphi$, $z$) with the unit vector ($\hat{\mathbf{r}}, \hat{\boldsymbol{\varphi}}, \hat{\mathbf{z}}$). As the magnetic skyrmion in a condensed-matter system, the probability density of single wave packet at a certain position gives the expected value of unit directional vector of the iSAM:

$$\mathbf{M} = \hbar\eta\left[\sin\Theta\cos\Phi\hat{\mathbf{r}} + \sin\Theta\sin\Phi\hat{\boldsymbol{\varphi}} + \mathrm{sgn}(l)\cos\Theta\hat{\mathbf{z}}\right], \tag{4}$$

where the unit directional vector **M** is determined by the sign of quantum number $l$, and $\eta$ is the envelope function; $\Theta$ and $\Phi$ are the polar angle and the azimuthal angle, respectively. Both the rotational symmetry of the optical system and the translational symmetry in the outer normal direction are considered here. Thus, the polar angle $\Theta$ is only radial-dependent and the azimuthal angle $\Phi$ is azimuthal-dependent. The **Equation 4** is logical because the geometry of photonic skyrmion (**Figure 1**) can be understood based on the expression used to calculate the skyrmion number for an arbitrary surface EM wave (see Supplemental Text V):

$$N_{\mathrm{SK}} = \frac{1}{4\pi}\iint_\Omega \mathbf{M}\cdot\left(\frac{\partial\mathbf{M}}{\partial x_i}\times\frac{\partial\mathbf{M}}{\partial x_j}\right)dx_i dx_j = \frac{1}{4\pi}\iint_\Omega \frac{1}{|\mathbf{\Sigma}|^3}\mathbf{\Sigma}\cdot\left(\frac{\partial\mathbf{\Sigma}}{\partial x_i}\times\frac{\partial\mathbf{\Sigma}}{\partial x_j}\right)dx_i dx_j. \tag{5}$$

where $\mathbf{M}=\mathbf{\Sigma}/|\mathbf{\Sigma}|$ denotes the unit directional vector of iSAM of the photonic skyrmion; and $x_i$ and $x_j$ represent two arbitrary orthogonal coordinates in the horizontal plane.

Obviously, the normalized quantity $|\mathbf{\Sigma}|$ can be extracted from the unit spin vectors **M**. The **Equation 5** reveals the bosonic feature of surface EM waves, which means that the statistical distributions of the surface EM fields do not affect the topological geometry of localized spin vectors, and thus the surface EM waves can be regarded as the quantized spin-1 bosons with population densities that are determined by the mode distributions or the field intensities. Furthermore, through **Figure 1**(a-b) and **Figure 1**(d-e) have different geometries, there is only one topology.[1] The topology can be characterized mathematically using an integer called the topological invariant, which is preserved under arbitrary continuous deformations. The photonic skyrmions are in the same topological phase and are topologically equivalent. Therefore, the chiral spin texture, in which the topological invariant can only change discretely, would not respond to continuous small perturbations. This topological invariant is in accordance with that of the electric field skyrmions.[45, 46] By projecting the winding of the spin vectors $\Sigma$ to a unit sphere, it can be determined that the topological skyrmion number is binary (±1) that is determined by sgn(*l*), as given in **Equation 4**.

The key point of the local distribution of the spin MD in spin-orbit coupling and the formation of the photonic skyrmion [as exhibited in **Equation 3**] is in line with the subluminal energy transport imposed by the relativity. The orbital MDs and spin MDs of the photonic skyrmions with vortex quantum number *l*=-2,3 can be found in **Figure 2**. Here, we provide an insight into the energy/momentum properties of photonic skyrmion in term of two aspects. On the one hand, the local energy-transport velocity (i.e. the group velocity in free space) can be determined using the relativistic relation between the energy W=⟨$\psi_0$|$\psi_0$⟩ and kinetic MD **p**: **p**=**v**$_g$W/$c^2$. Using the example of a light beam propagating in free space, it can be deduced that the energy density, the kinetic MD, the spin MD and the orbital MD of a plane EM wave propagating in the *z*-direction are W$^{pw}$=1/$\varepsilon_0$, **p**$^{pw}$= $v_g$W/$c^2$**ẑ**, **p**$_s^{pw}$=0 and **p**$_o^{pw}$=$v_{go}$W/$c^2$**ẑ**, respectively. Obviously, the spin MD does not transport energy for a light beam propagating in free space. In addition, the group velocity of this scalar wave is $v_g$=$v_{go}$=$c$ and cannot be superluminal. In constrast, for the photonic skyrmion, the orbital MD is related to the energy density as follows (see Supplemental Text II)

$$\mathbf{p}_o \approx \frac{l}{\omega r} W \hat{\boldsymbol{\varphi}} = \hat{\boldsymbol{\varphi}} \left( \frac{l}{k_0 r} c \right) W \Big/ c^2 , \tag{6}$$

with the group velocity of the canonical momentum $v_{go}$=($l$/$k_0 r$)$c$. If the spin MD vanishes in a manner similar to that of a light beam in free space, the group velocity $v_{go}$ can be greater than the velocity of light in vacuum in the region where $l$/$k_0 r$>1 as shown in the dotted circles in **Figure 2(d)** and **Figure 2(i)**. This seems illogical because the superluminal local energy-transport velocity contradicts the relativity theory.

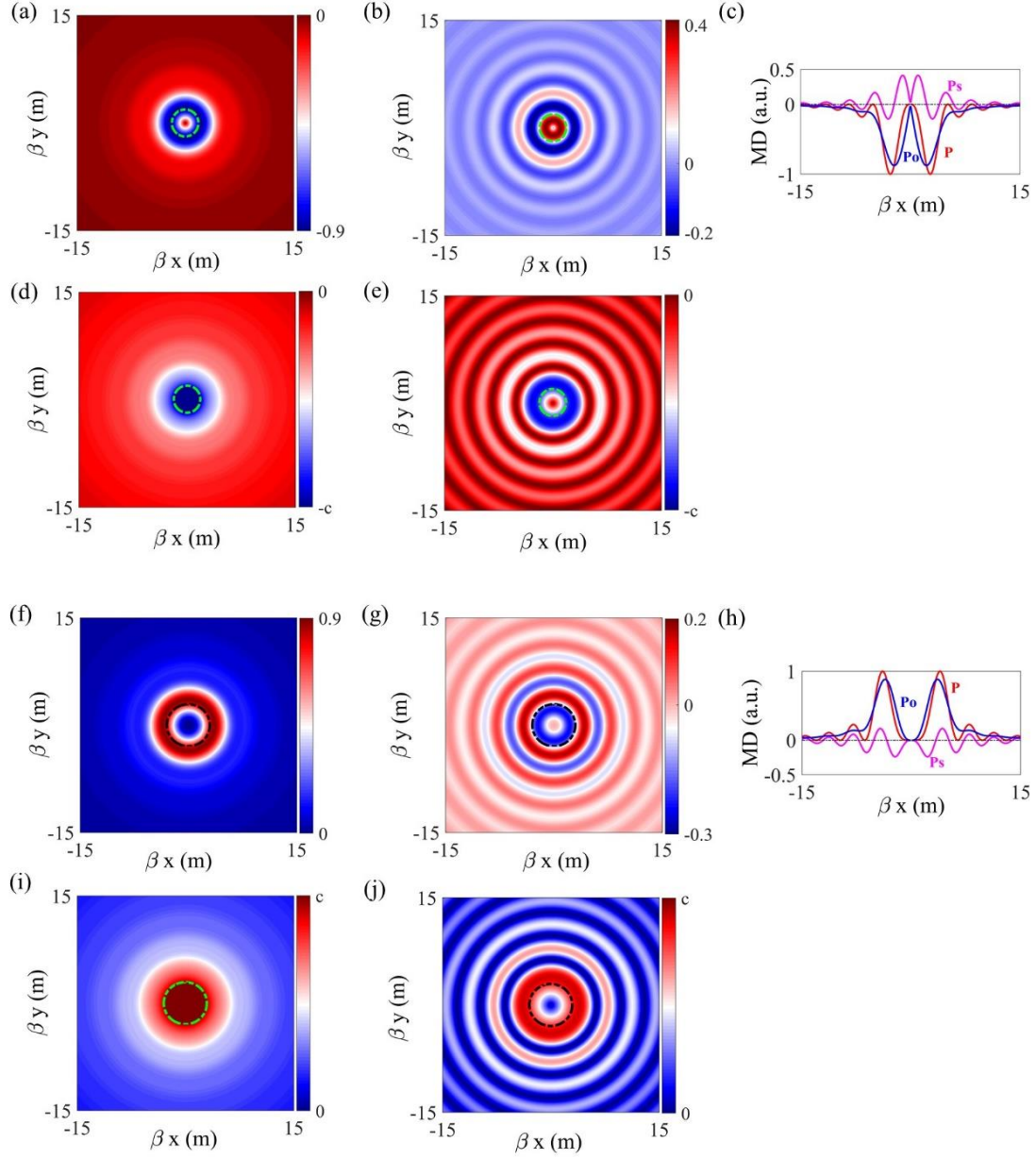

**Figure 2.** Dynamical properties of photonic skyrmions. (a) Orbital/canonical MD, (b) spin MD, (c) 1D contours of kinetic MD (red line), orbital/canonical MD (blue line), spin MD (magenta line), (d) orbital/canonical group velocity and (e) group velocity of photonic skyrmion with vortex topological charge $l=-2$; (f) orbital/canonical MD, (g) spin MD, (h) 1D contours of kinetic MD (red line), orbital/canonical MD (blue line), spin MD (magenta line), (i) orbital/canonical group velocity and (j) group velocity of photonic skyrmion with vortex topological charge $l=+3$. The dotted circles indicate the region where $l/k_0 r=1$. In the region where $l/k_0 r>1$, the spin MD must be inverted with respect to the orbital/canonical MD, which leads to the subluminal group velocity of the photonic skyrmion. The opearting wavelength is 633nm and the MDs are normalized with respect to the maximal absolute value of the corresponding kinetic MD. Here, the photonic skyrmions are constructed by the $p$-polarized light at air-gold interface.

On the other hand, as we proved in **Equation 2**, $\langle \mathbf{p}_s \rangle = 0$, and thus the integral spin MD does not carry net energy for the photonic skyrmion. This disappearance of the total spin MD is consistent with the transverse feature of the photonic skyrmion:[32] the longitudinal spin MD (either parallel or antiparallel to the energy transport direction) resultss in the transverse spin and the integration of this transvers spin vanishes because the transverse spin is independent of the helix or chirality of the light. However, as shown in **Figure 2**(b) and **Figure 2**(g), the local spin momentum definitely exists. In particular, from **Figure 2**(a-c) to **Figure 2**(f-h), it can be seen that the spin MDs are inverted with respect to the corresponding orbital MD and kinetic MD in the region where $l/k_0r>1$. The backward spin MDs reduce the total MD and the corresponding group velocities then become subluminal, as shown in **Figure 2**(e) and **Figure 2**(j). Thus, the local distribution of the spin MD, which can be either parallel or antiparallel, thus plays a critical role in both the spin-orbit coupling and the formation of the photonic skyrmion.

By substituting **Equation 4** into the variational **Equation 3**, we can obtain the 2nd-order differential equation $-2\partial^2\Theta/\partial r^2+2\sin2\Theta/r^2=0$, which can be solved numerically with the Runge-Kutta method or the Euler method (Supplemental Text VI). The numerical and theoretical results obtained for photonic skyrmions with quantum numbers of $l=\pm1$ and $l=\pm2$ are shown in **Figure 3**. Only the 1D contours are shown here because of the rotational symmetry of the spin textures. As theoretical predicted, at the point where the absolute value of $S_z$ is maximal, the $S_x$ disappears, thus indicating the spin vector is vertical. In constrast, in the case where $S_z$ is vanishing, the absolute value of $S_x$ reaches a maximum, thus indicating that the spin vector is oriented along the horizontal direction. There are two reasons for the slight distinction between the numerical and theoretical results here: first, **Equation 6** is approximate and becomes accurate at the large radius value; second, the local photon population will also affect the local profile of the chiral spin texture. However, the chiral spin textures, which vary from the 'up' state to the 'down' state for $l>0$ and from the 'down' state to the 'up' state for $l<0$, can definitely be observed. Therefore, the numerical results clearly verify the validity of our variational theory based on the conservation of total AM and spin-orbit coupling.

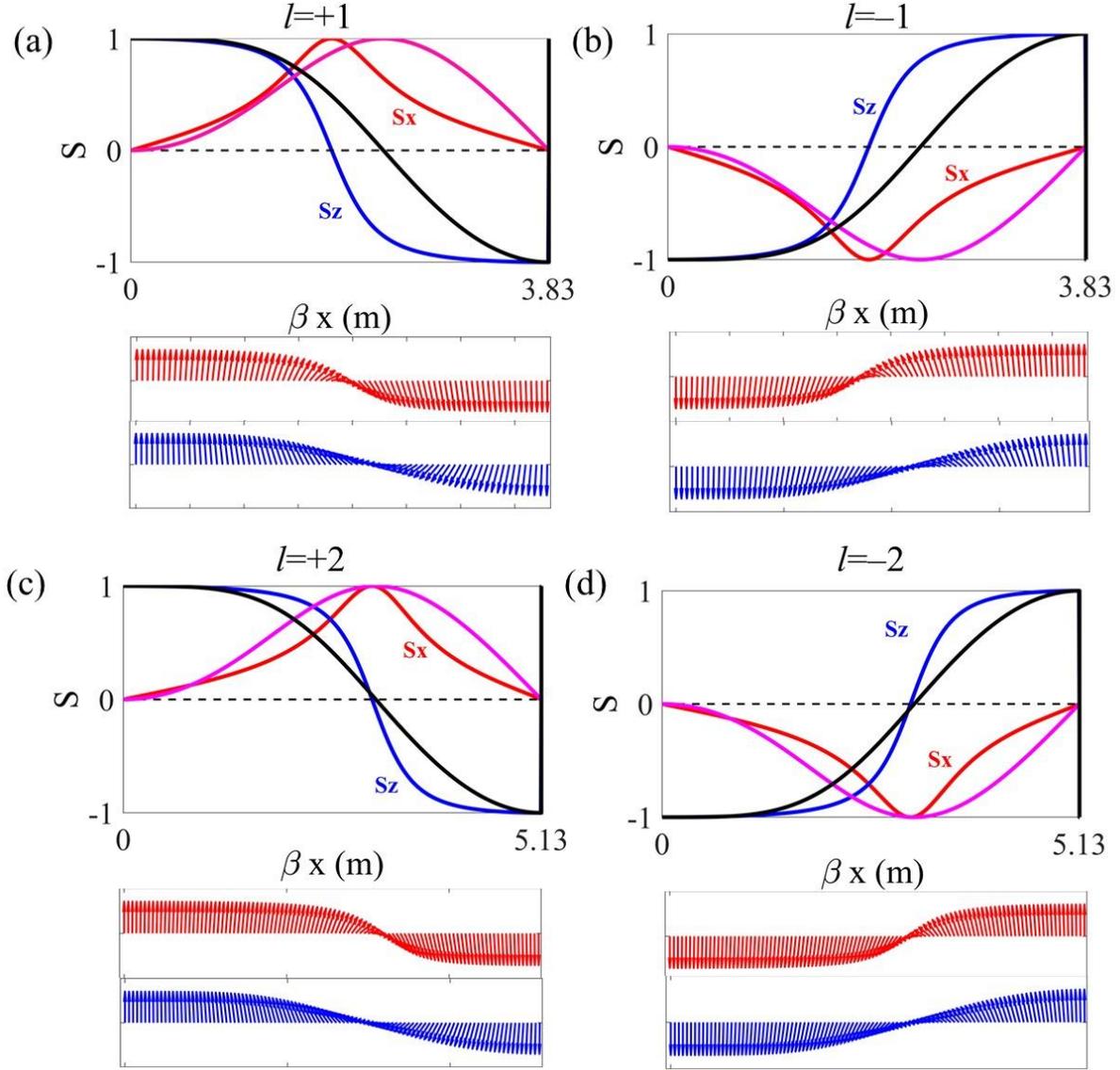

**Figure 3.** Comparison of theoretical and variational results for the skyrmion-like spin textures. 1D contours of $S_z$ and $S_x$ for the photonic skyrmions with quantum numbers (a) $l=+1$; (b) $l=-1$; (c) $l=+2$; and (d) $l=-2$. The $S_z$ and $S_x$ characteristics of the photonic skyrmions formed by Bessel-type surface waves are labelled by blue and red lines, respectively, while the corresponding $S_z$ and $S_x$ characteristics solved using the variational **Equation 3** are shown as black and magenta lines, respectively (top panels). The middle panels show the spin textures of the photonic skyrmions formed by Bessel-type surface waves with various quantum numbers. The bottom panels show the spin textures given by solving the variational **Equation 3**. The 2nd-order differential equation is solved using the Runge-Kutta method or Euler method and the boundary conditions are: $\Theta(0.001)=0$, $\Theta'(0.001)=1\times10^{-3}$ for $l=+1$ and $\Theta(0.001)=\pi$, $\Theta'(0.001)=-1\times10^{-3}$ for $l=-1$; $\Theta(0.001)=0$, $\Theta'(0.001)=5.5\times10^{-4}$ for $l=+2$ and $\Theta(0.001)=\pi$, $\Theta'(0.001)=-5.5\times10^{-4}$ for $l=-2$. Here, the photonic skyrmions are constructed by the *p*-polarized light at air-gold interface. The operating wavelength is 632.8nm. The horizontal axis of vectorgraph is consistent with the corresponding 1D contours of spin vectors.

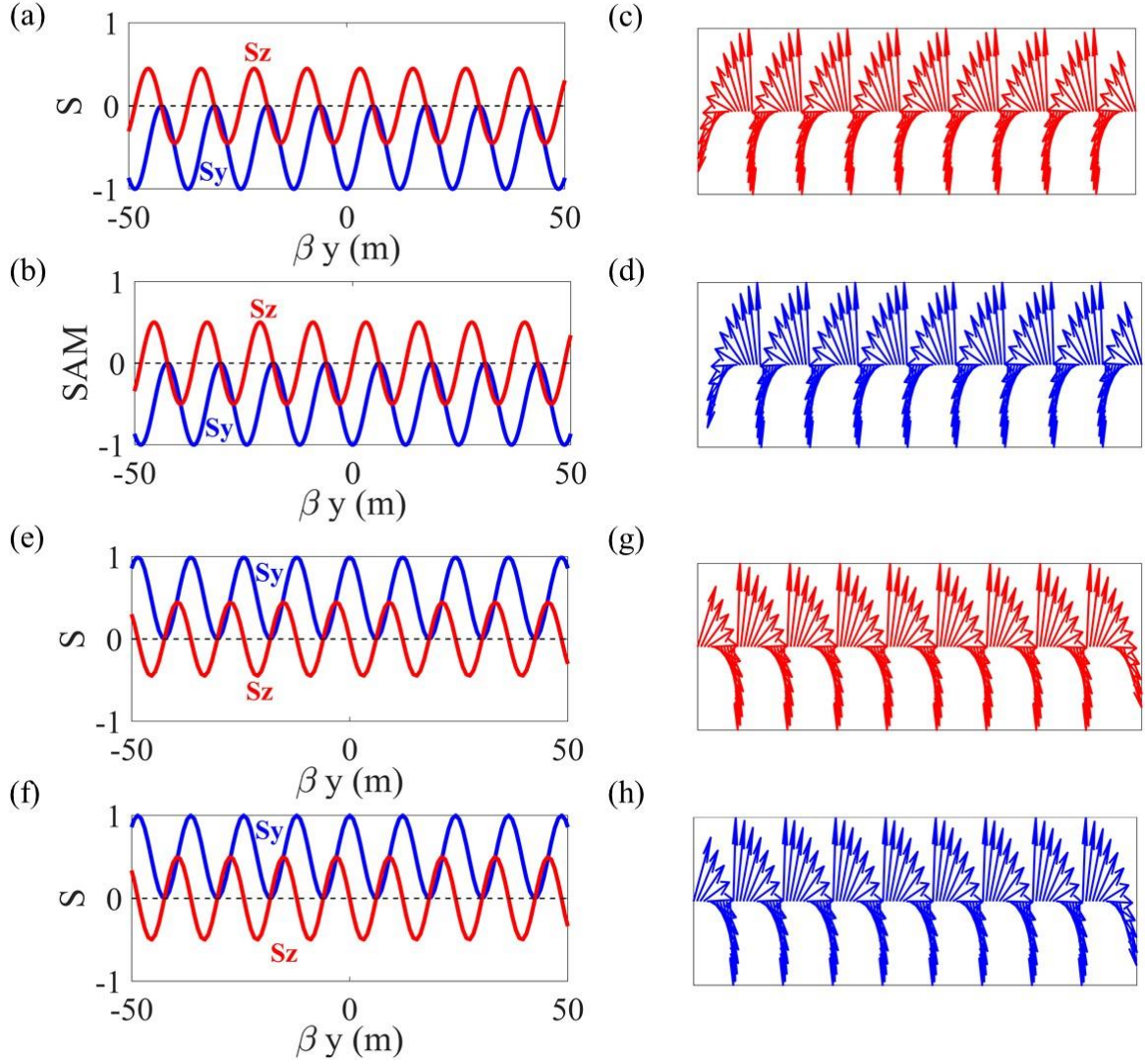

**Figure 4.** Comparison of domain wall-like spin texture between theoretical and variational results. (a) the theoretical and (b) variational results as the in-plane wave number component $k_y=k_r\sin(15°)$. The corresponding vector vectorgraphs for (c) theoretical (red) and (d) variational (blue) results in (a) and (b), respectively. (e) the theoretical and (f) variational results as the in-plane wave number component $k_y=k_r\sin(165°)$. The corresponding vector vectorgraphs for (g) theoretical (red) and (g) variational (blue) results in (e) and (f), respectively. The theoretical and variational results match very well. Here, the photonic domain wall-like textures are constructed by the *p*-polarized light at air-gold interface. The wavelength is 632.8nm. The horizontal axis of vectorgraph is consistent with the corresponding 1D contour of spin vectors.

In addition, , for the photonic chiral spin texture in Cartesian coordinate systems, the orbital MD is related to the energy density as follows (see Supplemental Text II)

$$\mathbf{p}_o^C = \left(ck_x/k_0\right)\hat{\mathbf{x}}\,\mathrm{W}^C/c^2 = v_{go}\hat{\mathbf{x}}\,\mathrm{W}^C/c^2, \qquad (7)$$

where the group velocity of canonical momentum is $v_{go}=(k_x/k_0)c$. Here, we assume the field propagates in *x*-direction and that $k_x$ is the in-plane wave number in propagating direction; the superscript *C* represents the Cartesian coordinate. Obviously, the group velocity $v_{go}$ can be greater than the velocity of light in vacuum at the region where $k_x/k_0>1$, as shown in **Figure S1**. In addition, as shown in **Equation S39** and **Equation S40**, it can found that $\langle \mathbf{p}_s \rangle=0$ and $\langle \mathbf{p}_o \rangle \neq 0$, which indicates the integral spin MD do not carry energy consistently. Using the variational **Equation 3**, we can obtain a 2nd-order differential equation $\partial^2 \Theta/\partial y^2=0$, which can be solved analytically (see Supplemental Text VI). The variational and theoretical results for photonic chiral spin textures in Cartesian coordinates are shown in **Figure 4**. It should be noted that the proportional relationship given by **Equation 7** is strictly valid in the Cartesian coordinate system. Therefore, we can see that the theoretical and variational results are wekk matched (and better matched than those in the cylindrical coordinate system), which thus demonstrates the validity of our theory.

Finally and most importantly, although this unified theory based on variational theorem can predict the magnetic and photonic chiral spin textures, as shown in **Figure 3** and **Figure 4**, respectively, there is a tremendous distinction between them: the chiral spin texture of the magnetic skyrmion is isolated, whereas the photonic chiral spin textures are approximatively periodic and are extended infinitely. This can be understood by considering the propagating wave nature of the wave packet, which indicates the photons are more suitable for the data transportation while the electrons are important for use in the data storage in optoelectronic applications.

## 3. Conclusion

In summary, we propose a novel framework based on the variational theory to reveal the formation and stability of the photonic chiral spin textures produced by the symmetry of an optical system. As demonstrated above, the formation of the chiral spin texture originates from conservation of the total AM and its stability is thus protected by the rotational symmetry of the optical system. Meanwhile, the chiral spin texture is only related to the wavelength via a propagating constant $\beta$. Thus, we can expect that the proposed theoretical framework will be suitable to be extended flexibly to EM waves with various wavelength and to use of various coordinate systems.[31] In addition, the research into spin-orbit interactions and spin-orbit decomposition has been extended into a variety of classical waves, including acoustic,[55] fluid,[56] elastic,[57] and even gravitational wave.[58] For example, longitudinal acoustic waves have a Klein-Gordon representation and there is a similar curl relationship between the momentum and the iSAM.[32, 59] This indicates that the idea of our theory can be extended to the classical wave field. Our findings are important for understanding of the chiral spin textures and spin dynamics of classical waves and for applications in manipulation (see Supplemental Text VII), imaging, nanometrology and prospective on-chip optoelectronic technologies.

**Supporting Information**

Supporting Information is available from the Wiley Online Library or from the author.

**Acknowledgements**

This work was supported, in part, by Guangdong Major Project of Basic Research No. 2020B0301030009, National Natural Science Foundation of China grants U1701661, 61935013, 62075139, 61427819, 61622504, and 61705135, leadership of Guangdong province program grant 00201505, Natural Science Foundation of Guangdong Province grant 2016A030312010, Science and Technology Innovation Commission of Shenzhen grants JCYJ20200109114018750, JCYJ20180507182035270, Shenzhen Peacock Plan KQTD20170330110444030. L.D. acknowledges the support given by the Guangdong Special Support Program.

**Conflict of Interest**

The authors declare no conflict of interest.